\documentclass[11pt,fleqn]{article}
\usepackage{amsmath,amssymb,amsthm,epsfig,graphics}

\newtheorem*{lemma*}{Lemma}

\newtheorem*{conjecture*}{Conjecture}

{\theoremstyle{definition}

\newtheorem{example}{Example}
\newtheorem{remark}{Remark}

\newtheorem*{note*}{Note}
}

\allowdisplaybreaks

\begin{document}

\begin{center}
\LARGE \bf
$S$-expansions of three-dimensional Lie algebras
\end{center}

\begin{center} \bf
Maryna NESTERENKO
\end{center}

\noindent Institute of Mathematics of NAS of Ukraine, 3
Tereshchenkivs'ka Str., Kyiv-4, 01601 Ukraine\\
E-mail: maryna@imath.kiev.ua


\begin{abstract}
\noindent
$S$-expansions of three-dimensional real Lie algebras are considered.
It is shown that the expansion operation allows one to obtain a non-unimodular Lie algebra from a unimodular one.
Nevertheless $S$-expansions define no ordering on the variety of Lie algebras of a fixed dimension.
\end{abstract}

\section{Introduction}\label{Introduction}
In 1961  Zaitsev~\cite{Nesteerenko:Zaitsev1969} supposed that all solvable Lie algebras of a fixed dimension could be obtained via contractions
from the semisimple algebras of the same dimension.
He called such solvable Lie algebras `limiting solvable' and proposed the contraction procedure for the `limiting classification' of solvable Lie algebras.
Later the same conjecture was also formulated by other scientists, e.g. Celeghini and Tarlini~\cite{Nesterenko:Celeghini&Tarlini1981}.
Complexity of the actual state of affairs was illustrated in~\cite{Nesterenko:NesterenkoPopovych2006}, where contractions of real and complex low-dimensional Lie algebras were studied.
The incorrectness of the above conjecture was illustrated by the fact that all semisimple (and reductive) Lie algebras are unimodular and
any continuous contraction of a unimodular algebra necessarily results in a unimodular algebra. Moreover, semisimple Lie algebras exist not in all dimensions.

The aim of this paper is to revisit Zaitsev's conjecture in terms of $S$-expansions.
We will study $S$-expansions of the real three-dimensional Lie algebras.
It was shown in~\cite{Nesterenko:Izaurieta2006} that generalized In\"on\"u--Wigner contractions give a particular case of $S$-expansions.
This is why we consider only pairs of such algebras that are not connected by a contraction.

The paper is arranged as follows.
Section 2 contains preliminary information on $S$-expansions and relevant objects.
In Section 3 we construct several key examples of three-dimensional $S$-expansions
and discuss the possibility of application of $S$-expansions to the classification of solvable Lie algebras.

\section{Basic properties of $S$-expansions}

Roughly speaking, the \emph{$S$-expansion} 
is the ``product'' of the semigroup and the Lie algebra
with a Lie algebra structure defined in a special way.
The notion was introduced in~\cite{Nesterenko:Izaurieta2006}.
It generalizes the notions of expansion, deformation (under a proper choice of the corresponding semigroup), extension
and generalized In\"on\"u--Wigner contraction.

\looseness=-1
Let $S=\{\lambda_\alpha\}$ be an Abelian semigroup of order $N$
and let $\mathfrak g$ be an $n$-dimensional Lie algebra with the structure constants $C_{ij}^k$
in a fixed basis $\{e_i\}$ of the underlying vector space~$V$.
Here and in what follows
the indices $\alpha$, $\beta$ and $\gamma$ run from~1 to~$N$ and
the indices $i$, $j$ and $k$ run from~1 to~$n=\dim V$.
The~$nN$-dimensional $S$-\emph{expanded Lie algebra} $\mathfrak G :=S\times\mathfrak g$,
the underlying vector space of which is spanned by the basis elements $e_{(i,\alpha)}=\lambda_\alpha e_i$,
is defined by the structure constants
\begin{equation}\label{Nesterenko:Cijk}
C_{(i,\alpha)(j,\beta)}^{(k,\gamma)}=\left\{
\begin{array}{ll}
C_{ij}^k \quad&\mbox{if} \quad \lambda_\alpha\lambda_\beta=\lambda_\gamma,\\
0\quad &\mbox{otherwise}.
\end{array}
\right.
\end{equation}

\begin{remark}\label{Nesterenko:basis change}
It directly follows from~(\ref{Nesterenko:Cijk}) that
the product of elements $e_{(i,\alpha)}$ and $e_{(j,\beta)}$ in the algebra $\mathfrak G$ is zero if $[e_i,e_j]=0$.
Therefore, an appropriate representative from the isomorphism class of the initial Lie algebra $\mathfrak g$ should be chosen if
$S$-expansion is used for a certain purpose.
In other words, sometimes the structure constants tensor has to be transformed
by an element of the group ${\rm GL}(V)$ before the expansion procedure.
Of course, commutation relations of the expanded algebra $\mathfrak G$ can also be reduced to an appropriate form using a change of basis.
\end{remark}

Let a Lie algebra $\mathfrak{g}$ admit the decomposition $\mathfrak{g}=\check{V}\oplus \hat{V}$ with $[\check{V},\hat{V}]\subset \hat{V}$, namely
\begin{equation}
[\check{v}_{\check{i}},\hat{v}_{\hat{j}}]=\hat{C}_{\check{i}\hat{j}}^{\hat{k}} \hat{v}_{\hat{k}},
\quad
[\check{v}_{\check{i}},\check{v}_{\check{j}}]=\check{C}_{{\check{i}}{\check{j}}}^{\check{k}} \check{v}_{\check{k}}+\hat{C}_{{\check{i}}{\check{j}}}^{\hat{k}}\hat{v}_{\hat{k}},
\quad
[\hat{v}_{\hat{i}},\hat{v}_{\hat{j}}]=\check{C}_{\hat{i}\hat{j}}^{\check{k}} \check{v}_{\check{k}}+\hat{C}_{\hat{i}\hat{j}}^{\hat{k}} \hat{v}_{\hat{k}};
\end{equation}
where $\{\check{v}_{\check{i}}\}$ is a basis of $\check{V}$ and $\{\hat{v}_{\hat{i}}\}$ is a basis of $\hat{V}$.
Then the elements $\{\check{v}_{\check{i}}\}$ form a well-defined \textit{reduced} Lie algebra $|\check{V}|$ with the Lie product defined by the structure constants $\check{C}_{\check{i}\check{j}}^{\check{k}}$.

\subsection{Algebras of lower dimensions}
In general case the dimension of the $S$-expanded Lie algebra~$\mathfrak G$ is greater than~$n$
and it has a more complicated structure than~$\mathfrak g$.
Therefore for some purposes it is reasonable to consider certain subalgebras or reduced algebras of less dimensions
instead of working with the whole expanded algebra.

The problem of the construction of algebras of less dimension that are related to an expanded Lie algebra is rather complicated.
Two classes of such algebras were presented in~\cite{Nesterenko:Izaurieta2006}
for $S$-expanded Lie algebras admitting an agreed decomposition of the associated semigroups and Lie algebras.
A semigroup $S$ and a Lie algebra $\mathfrak g$ are said to admit \emph{a resonant subset decomposition}
if they can be decomposed as
$\mathfrak g=\bigoplus_{p\in I} V_p$ (where the direct sum is interpreted in the sense of vector spaces only)
and $S=\bigcup_{p\in I}S_p$ with some index set~$I$ in such a way that
$[V_p,V_q]\subset\bigoplus_{r\in i(p,q)}V_r$ and $S_p S_q\subset\bigcap_{r\in i(p,q)}S_r$,
where for every $p,q\in I$ the index set $i(p,q)$ is a subset of~$I$.
In the resonance case,
$\mathfrak G_R=\bigoplus_{p\in I} (S_p\times V_p)$ is a \emph{resonance subalgebra} of the $S$-expanded Lie algebra~$\mathfrak G$.
For each resonant decomposition this gives a Lie algebra of dimension less than the dimension of~$\mathfrak G$.
Algebras of the other class are constructed by means of the reduction of Lie algebras.
Suppose that for each $p\in I$ the set $S_p$ is partitioned into the subsets $\check S_p$ and~$\hat S_p$
such that $\smash{\check S_p \hat S_q\subset\bigcap_{r\in i(p,q)}\hat S_r}$.
This partition induces the decomposition $\smash{\mathfrak G_R=\hat{\mathfrak G}\oplus\check{\mathfrak G}}$,
where direct sums are interpreted in the sense of vector spaces,
$\smash{\check{\mathfrak G}=\bigoplus_{p\in I} (\check S_p\times V_p)}$ and $\smash{\hat{\mathfrak G}=\bigoplus_{p\in I} (\hat S_p\times V_p)}$.
As $\smash{[\check{\mathfrak G},\hat{\mathfrak G}]\subset\hat{\mathfrak G}}$,
the projection of the Lie bracket of $\mathfrak G_R$ to $\check{\mathfrak G}$ gives a well-defined Lie bracket on $|\check{\mathfrak G}|$.
In other words, the Lie algebra $\check{\mathfrak G}$ with this bracket is the reduced algebra for~$\mathfrak G_R$.

Given an Abelian semigroup~$S$ with a zero $\lambda_N=0_S \in S$, i.e., $\lambda_\alpha 0_S=0_S$ for all $\alpha=1,\dots, N$,
any $S$-expanded Lie algebra $\mathfrak{G}=S\times \mathfrak{g}$ can be decomposed as
$\mathfrak{G}=\check{V}\oplus \hat{V}$ with $[\check{V},\hat{V}]\subset \hat{V}$,
where $\hat{V}=\langle 0_Se_1,0_Se_2,\dots, 0_Se_n \rangle$.
In this case the reduced Lie algebra $\mathfrak{G}_0:=|\check{V|}$ 
is called the \textit{$0_S$-reduced algebra} of $\mathfrak{G}$.

\begin{remark}\label{Nesterenko:right_side}
The condition~(\ref{Nesterenko:Cijk}) implies that the $0_S$-reduced algebra $\mathfrak{G}_0$ has the following commutation relations
in the basis  $\{\lambda_\alpha e_i\mid\alpha=1,\dots, N-1,\ i=1,\dots,n\}$:
\begin{equation}\label{Nesterenko:ei_ej}
[\lambda_\alpha e_i,\lambda_\beta e_j]=
\left\{
\begin{array}{ll}
\sum\limits_{k=1}^nC_{ij}^k \lambda_\alpha\lambda_\beta e_k \quad &\text{if} \quad \lambda_\alpha\lambda_\beta\ne 0_S,\\
0\quad &\text{if} \quad \lambda_\alpha\lambda_\beta=0_S.
\end{array}
\right.
\end{equation}
\end{remark}

\subsection{$S$-expansions and contractions}

As generalized In\"on\"u--Wigner contractions are related to gradings of the contracted Lie algebras
(i.e., to special decompositions of the underlying vector space),
it is clear that all generalized IW-contractions can be obtained by means of resonant $S$-expansions.
See~\cite{Nesterenko:Izaurieta2006} for details.

At the same time, this claim does not imply that $S$-expansions exhaust all possible contractions
since there exist contractions that are not realized by generalized IW-contractions~\cite{Nesterenko:Burde2005,Nesteerenko:Popovych2010}.

\begin{example}
Consider the example of such a contraction constructed in~\cite{Nesterenko:Burde2005} for the seven-dimensional Lie algebras~$\mathfrak g_F$ and~$\mathfrak g_E$.
These algebras are defined by the commutation relations
\begin{gather*}
\mathfrak g_F\colon [e_1,e_i]=e_{i+1}, \, 2\leqslant i \leqslant 6,\\
\phantom{\mathfrak g_F\colon }
[e_2,e_3]=e_6,
\quad
[e_2,e_4]=e_7,
\quad
[e_2,e_5]=e_7,
\quad
[e_3,e_4]=-e_7,
\\
\mathfrak g_E\colon
[e_1,e_i]=e_{i+1}, \, 2\leqslant i \leqslant 6,
\quad
[e_2,e_3]=e_6+e_7,
\quad
[e_2,e_4]=e_7
\end{gather*}
and are characteristically nilpotent since their differentiation algebras
\begin{gather*}
{\rm Der}(\mathfrak g_F)=\langle
2E_{21}+E_{42}+E_{53}+3E_{64}+5E_{75}+2E_{76},\
E_{31}+E_{52}-E_{75},\
\\
\phantom{{\rm Der}(\mathfrak g_F)=\langle}
E_{32}+E_{43}+E_{54}+E_{65}+E_{76},\
E_{61},\
E_{41}-E_{51}+E_{74},\ 
E_{51}+E_{62}
\rangle,
\end{gather*}
\begin{gather*}
{\rm Der}(\mathfrak g_E)=\langle
E_{31}-E_{41}+E_{52},\
E_{41}+E_{62},\
-E_{31}+E_{41}+E_{63}+E_{74},
\\
\phantom{{\rm Der}(\mathfrak g_E)=\langle}
-E_{21}-E_{31}+E_{41}+2E_{42}+2E_{53}+E_{63}+E_{64},\
E_{71},\
E_{51},\
\\
\phantom{{\rm Der}(\mathfrak g_E)=\langle}
E_{21}+E_{31}-E_{41}-E_{42}-E_{53}-E_{63}+E_{75},\
-E_{41}+E_{73},
\\
\phantom{{\rm Der}(\mathfrak g_E)=\langle}
E_{32}+E_{43}+E_{54}+E_{65}+E_{76},\
E_{72},\
E_{61}
\rangle
\end{gather*}
are nilpotent.
Here $E_{ij}$ denotes the $7 \times 7$ matrix with only nonzero $(i,j)$th entry
that equals~1.
(It is obvious that both the differentiation algebras consist of lower triangular matrices.)
This means that the algebras~$\mathfrak g_F$ and~$\mathfrak g_E$ admit no grading
and hence none of them can be a result of a generalized IW-contraction.
On the other hand, there exist the contraction from~$\mathfrak g_F$ to~$\mathfrak g_E$
provided by the following contraction matrix at $\varepsilon\to 0$:
\begin{gather*}
U_\varepsilon=\left(
\begin{array}{ccccccc}
\varepsilon & 0 & 0 & 0 & 0 & 0 & 0\\
0 & \varepsilon^3 & 0 & 0 & 0 & 0 & 0\\
0 & 0 & \varepsilon^4 & 0 & 0 & 0 & 0\\
0 & \tfrac12\varepsilon^4 & 0 & \varepsilon^5 & 0 & 0 & 0\\
0 & 0 & \tfrac12\varepsilon^5 & 0 & \varepsilon^6 & 0 & 0\\
0 & 0 & 0 & \tfrac12\varepsilon^6 & 0 & \varepsilon^7 & 0\\
0 & 0 & 0 & 0 & \tfrac12\varepsilon^7 & 0 & \varepsilon^8
\end{array}
\right).
\end{gather*}
\end{example}

Examples on non-universality of the generalized IW-contractions in dimension four can be found in~\cite{Nesteerenko:Popovych2010}.
This is why it is still unclear whether all contractions can be obtained via $S$-expansions.

On the other part, there exist $S$-expansions which are not equivalent to contractions.
See, e.g., the example on a connection between the Lie algebras $\rm{sl}(2,\mathbb{R})$ and $A_{2.1}\oplus A_1$ that is presented in the next section.

Other objects closely related to $S$-expansions are given by the purely algebraic notion of \emph{graded contractions}~\cite{Nesterenko:Patera1991}.
The graded contraction procedure is the following. Structure constants of a graded Lie algebra are multiplied by numbers which
are chosen in such a way that the multiplied structure constants define a Lie algebra with the same grading.
Graded contractions include discrete contractions as a subcase but do not cover all continuous~ones.


Contractions of real and complex three-dimensional Lie algebras were completely studied in~\cite{Nesterenko:NesterenkoPopovych2006}.
As all these contractions are realized by the generalized In\"on\"u--Wigner contractions, they can be realized by $S$-expansions.
Moreover, \mbox{$S$-expansions} lead to establishing new relations between such algebras.


\section{Three-dimensional $S$-expansions}
According to Mu\-ba\-rak\-zya\-nov's classification~\cite{Nesterenko:Mubarakzyanov1963},
nonisomorphic real three-dimen\-sional Lie algebras are exhausted by
the two simple algebras $\rm{sl}(2,\mathbb{R})$ and $\rm{so}(3)$;
the two parameterized series of solvable algebras $A_{3.4}^a,\ 0{<}|a|{\leqslant}1$ and $A_{3.5}^b,\ b{\geqslant}0$;
the three single indecomposable solvable algebras $A_{3.1}$, $A_{3.2}$ and $A_{3.3}$; and the two decomposable solvable algebras $3A_1$ and $A_{2.1}\oplus A_1$.

In~\cite{Nesterenko:NesterenkoPopovych2006} it was proven
that all unimodular three-dimensional algebras (namely, $\rm{sl}(2,\mathbb R)$, $\rm{so}(3)$, $A_{3.4}^{-1}$, $A_{3.5}^0$, $A_{3.1}$ and $3A_1$)
belongs to the orbit closure of at least one of the simple algebras.

\begin{figure}[ht]
\centerline{\includegraphics[scale=1.]{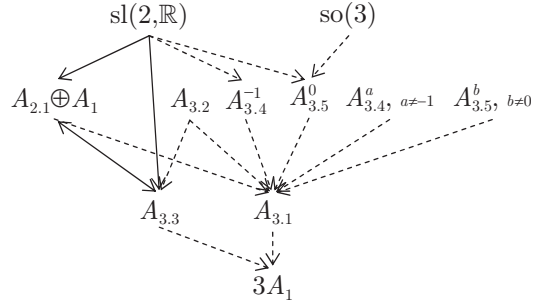}}
\caption{Contractions of real three-dimensional Lie algebras are marked by the dashed lines
and $S$-expansions which are not equivalent to contractions are marked by solid lines.}\label{nesterenko:fig1}
\end{figure}

All contractions are shown on Figure~\ref{nesterenko:fig1} by dashed lines (each arrow indicates the direction of the corresponding contraction).
Repeated contractions (i.e., contractions of the kind: $\mathfrak g$ contracts to $\mathfrak g_0$ if $\mathfrak g$ contracts to $\mathfrak g_1$ and $\mathfrak g_1$ contracts to $\mathfrak g_0$)
are implied but not indicated on this figure.
All three-dimensional contractions were realized by generalized IW-contractions, therefore the corresponding $S$-expansions are also exist.

There are a number of necessary conditions to be satisfied for a pair of Lie algebras connected by a contraction, see, e.g.,~\cite{Nesterenko:NesterenkoPopovych2006}.
The following examples show that $S$-expansions of Lie algebras do not obey the major part of these rules.
These examples involve the algebras (for each algebra we present only nonzero commutation ralations)
\begin{gather}\label{Nesterenko:sl2R}
\rm{sl}(2,\mathbb{R})\colon\quad [e_1,e_2]=e_1,\ [e_2,e_3]=e_3,\ [e_1,e_3]=2e_2;
\\
\label{Nesterenko:A21}
A_{2.1}\oplus A_1\colon\quad [e_1,e_2]=e_1;
\\
\label{Nesterenko:A33}
A_{3.3}\colon\quad [e_1,e_3]=e_1,\ [e_2,e_3]=e_2.
\end{gather}

\subsection{Unimodularity of $S$-expansions}

Given a unimodular Lie algebra $\mathfrak g\langle e_1,\dots, e_n\rangle$, 
we have
\begin{equation}\label{nesterenko:tr_ad}
{\rm tr}({\rm ad}_{e_i})=\sum_{j=1}^nC_{ij}^j=0\qquad \forall\; i=1,\dots, n.
\end{equation}

Using~(\ref{nesterenko:tr_ad}) for the basis elements $E_{(i-1)N+\alpha}:=\lambda_\alpha e_i$ of the $S$-expanded Lie algebra $\mathfrak{G}$, we get
\[
{\rm tr}({\rm ad}_{E_{(i-1)N+\alpha}})=
\sum_{\beta=1}^N\sum_{j=1}^n K_{\alpha\beta}^\beta C_{ij}^j=
\sum_{\beta=1}^N K_{\alpha\beta}^\beta\Bigg(\sum_{j=1}^n C_{ij}^j\Bigg)=0,
\]
where
\[K_{\alpha\beta}^\gamma= \left\{\begin{array}{l}
1, \quad \text{if} \quad \lambda_\alpha\lambda_\beta=\lambda_\gamma,\\
0,\quad \text{otherwise}.
\end{array}
\right.
\]
Thereby the unimodularity property is necessarily preserved by any sole $S$-expansion.
At the same time, a unimodular Lie algebra may contain a non-unimodular subalgebra
or may be reduced to a non-unimodular Lie algebra.

\subsection{Examples of non-unimodular $S$-expansions}
Consider the expansion of the Lie algebra $\rm{sl}(2,\mathbb{R})$ by means of the Abelian semigroup \mbox{$S_2:=\{\lambda_1,\lambda_2\}$}
where $\lambda_1\lambda_1=\lambda_1\lambda_2=\lambda_2\lambda_1=\lambda_2\lambda_2=\lambda_2$,
i.e., $\lambda_2$ is the zero element $0_{S_2}$ of the semigroup.
The elements $E_{2(i-1)+\alpha}:=\lambda_\alpha e_i$, $\alpha=1,2$, $i=1,2,3$, form a basis of six-dimensional $S_2$-expanded Lie algebra $\mathfrak{G}$.
Then from~(\ref{Nesterenko:sl2R}) we obtain the nonzero commutation relations
\[
\begin{tabular}{llll}
$[E_1,E_3]=E_2$,&
$[E_1,E_4]=E_2$,&
$[E_1,E_5]=2E_4$,&
$[E_1,E_6]=2E_4$,\\[.7ex]
$[E_2,E_3]=E_2$,&
$[E_2,E_4]=E_2$,&
$[E_2,E_5]=2E_4$,&
$[E_2,E_6]=2E_4$,\\[.7ex]
$[E_3,E_5]=E_6$,&
$[E_3,E_6]=E_6$,&
$[E_4,E_5]=E_6$,&
$[E_4,E_6]=E_6$.
\end{tabular}
\]

Elements $E_1$, $E_2$ and $E_3$ span a three-dimensional subalgebra with the nonzero commutation relations $[E_1,E_3]=E_2$, $[E_2,E_3]=E_2$.
The basis change $\tilde{E}_1=E_2$, $\tilde{E}_2=E_3$, $\tilde{E}_3=E_1-E_2$ leads to the unique nonzero relation $[\tilde{E}_1,\tilde{E}_2]=\tilde{E}_1$ and,
therefore, we obtain the Lie algebra $A_{2.1}\oplus A_1$, c.f.~(\ref{Nesterenko:A21}).

Another example concerns the Lie algebras $\rm{sl}(2,\mathbb{R})$ and $A_{3.3}$.
Let a three-ele\-ment Abelian semigroup with a zero $S_3=\{0_{S_3}=\lambda_1,\lambda_2,\lambda_3\}$ satisfies the conditions
$\lambda_2\lambda_3=\lambda_3\lambda_2=\lambda_2$ and $\lambda_2\lambda_2=\lambda_3\lambda_3=\lambda_1$.
The above properties of~$S_3$ are consistent with the semigroup structure and are sufficient for defining $S_3$-expansions.
Thus, the $S_3$-expansion of $\rm{sl}(2,\mathbb{R})$ is the nine-dimensional Lie algebra $S_3\times \rm{sl}(2,\mathbb{R})$ with the nonzero commutation relations
\[
\begin{tabular}{lll}
$[E_1,E_4]=E_1$,&
$[E_1,E_5]=E_1$,&
$[E_1,E_6]=E_1$,
\\[.7ex]
$[E_1,E_7]=2E_4$,&
$[E_1,E_8]=2E_4$,&
$[E_1,E_9]=2E_4$,
\\[.7ex]
$[E_2,E_4]=E_1$,&
$[E_2,E_5]=E_1$,&
$[E_2,E_6]=E_2$,
\\[.7ex]
$[E_2,E_7]=2E_4$,&
$[E_2,E_8]=2E_4$,&
$[E_2,E_9]=2E_5$,
\\[.7ex]
$[E_3,E_4]=E_1$,&
$[E_3,E_5]=E_2$,&
$[E_3,E_6]=E_1$,
\\[.7ex]
$[E_3,E_7]=2E_4$,&
$[E_3,E_8]=2E_5$,&
$[E_3,E_9]=2E_4$,
\\[.7ex]
$[E_4,E_7]=E_7$,&
$[E_4,E_8]=E_7$,&
$[E_4,E_9]=E_7$,
\\[.7ex]
$[E_5,E_7]=E_7$,&
$[E_5,E_8]=E_7$,&
$[E_5,E_9]=E_8$,
\\[.7ex]
$[E_6,E_7]=E_7$,&
$[E_6,E_8]=E_8$,&
$[E_6,E_9]=E_7$.
\end{tabular}
\]

From the algebra $S_3\times \rm{sl}(2,\mathbb{R})$ we can extract the three-dimensional subalgebra $\langle E_1, E_2, E_6\rangle$
isomorphic to $A_{3.3}$, c.f.~(\ref{Nesterenko:A33}).

The two above constructions  give us examples on connection between unimodular and non-unimodular three-dimensional Lie algebras
by means of the expansion and subalgebra extraction.

\begin{remark}
Concerning the rest of non-unimodular Lie algebras, namely $A_{3.2}$, $A_{3.4}^a$ and $A_{3.5}^b$,
it seems to be impossible to construct them from simple three-dimensional Lie algebras by means of $S$-expansion.
This conjecture is motivated by the disagreement of the right-hand sides of the respective canonical commutators
that can not be overcome by basis changes. Nevertheless, this conjecture needs a rigorous proof.
\end{remark}

\subsection{$S$-expansion from $A_{2.1}\oplus A_1$ to $A_{3.3}$ and vice versa}
Consider $S_3$-expansion of the Lie algebra $A_{2.1}\oplus A_1$.
To skip the tedious commutation relations of the nine-dimensional Lie algebra we consider only those which concern the basis elements
$E_1=\lambda_1e_1$, $E_2=\lambda_2e_1$ and $E_6=\lambda_3e_2$.
They are
\[[E_1,E_6]=E_1,\; [E_2,E_6]=E_2,\; [E_1,E_2]=0.\]
This implies that the basis elements $E_1$, $E_2$ and $E_6$ form a subalgebra isomorphic to $A_{3.3}$, c.f.~(\ref{Nesterenko:A33}).

The inverse connection can be obtained by means of $S_2$-expansion of the algebra~$A_{3.3}$.
The basis elements $E_1$, $E_2$ and $E_6$ span the subalgebra isomorphic to $A_{2.1}\oplus A_1$.
Indeed, the nontrivial commutation relations between these elements are
\[[E_1,E_2]=0,\quad [E_1,E_6]=E_1,\quad [E_2,E_6]=E_1.\]
After the basis change $\tilde{E}_1=E_1-E_2$, $\tilde{E}_2=E_6$, $\tilde{E}_3=E_2$,
we obtain the unique nonzero commutation relation $[\tilde{E}_1,\tilde{E}_2]=\tilde{E}_1$.

In contrast to the contraction procedure,
the last expansion creates a Lie algebra of more complicated structure.
For example,
the dimension of the center decreases from 1 to 0,
the dimension of the Cartan subalgebra decreases from 2 to 1
and the dimension of the derivative enlarges from 1 to 2
after the $S$-expansion from the Lie algebra $A_{2.1}\oplus A_1$ to the algebra $A_{3.3}$.

At the same time, the $S$-expansion, even combined with algebra reduction and singling out a subalgebra, preserves
certain properties of Lie algebras.
Thus, we have $[S\times\mathfrak g,S\times\mathfrak g]=(SS)\times[\mathfrak g,\mathfrak g]$.
Hence the algebra $S\times\mathfrak g$ is solvable (resp.\ nilpotent) if and only if the algebra $\mathfrak g$ is,
and then the solvability (resp.\ nilpotency) degrees of these algebras coincide.
The procedures of reducing the algebra $S\times\mathfrak g$ and singling out a subalgebra
may not increase the solvability (resp.\ nilpotency) degree.

Note that all four of the discussed examples of expansions
can be obtained from the non-Abelian two-dimensional Lie algebra,
since in all cases the key role is played by the commutation relation $[e_1,e_2]=e_1$.

\begin{remark}
$S$-expansions do not set any ordering relationship on the variety of Lie algebras of a fixed dimension.
This statement is illustrated by the example of Lie algebras $A_{2.1}\oplus A_1$ and $A_{3.3}$ that can be connected by $S$-expansion in both directions.
Therefore, in spite of the fact that it is possible to construct non-unimodular Lie algebras from unimodular ones,
there is still a question whether $S$-expansions fit to the classification of solvable Lie algebras of a fixed dimension by means of $S$-expansions of simple (semisimple) Lie algebras of the same dimension.
\end{remark}

\subsection*{Acknowledgements}

The author is grateful to Professor Roman O.~Popovych for helpful discussions and useful censorious remarks.


\begin{thebibliography}{99}
\footnotesize
\itemsep0mm

\bibitem{Nesterenko:Burde2005}
Burde D.,
Degenerations of 7-dimensional nilpotent Lie algebras,
\textit{Comm. Algebra} {\bf 33} (2005), 1259--1277.

\bibitem{Nesterenko:Celeghini&Tarlini1981}
Celeghini E. and Tarlini M.,
Contraction of group representations II,
\textit{Nuovo Cimento B} \textbf{65} (1981), 172--180.

\bibitem{Nesterenko:Patera1991}
Couture M., Patera J., Sharp R.T. and Winternitz P.,
Graded contractions of $sl(3,\mathbb{C})$,
\textit{J.~Math. Phys.} \textbf{32} (1991), 2310--2318.

\bibitem{Nesterenko:Izaurieta2006}
Izaurieta F., Rodriguez E. and Patricio Salgado P.,
Expanding Lie (super)algebras through Abelian semigroups,
\textit{J. Math. Phys.} {\bf 47} (2006), 123512.

\bibitem{Nesterenko:Mubarakzyanov1963}
Mubarakzyanov~G.M.,
On solvable Lie algebras,
\textit{Izv. Vys. Ucheb. Zaved. Matematika}, \textbf{1(32)} (1963), 114--123 (Russian).

\bibitem{Nesterenko:NesterenkoPopovych2006}
Nesterenko M. and Popovych R.,
Contractions of low-dimensional Lie algebras,
\textit{J. Math. Phys.} {\bf 47} (2006), 123515.

\bibitem{Nesteerenko:Popovych2010}
Popovych D.R. and Popovych R.O.,
Lowest dimensional example on non-universality of generalized In\"on\"u--Wigner contractions,
\textit{J. Algebra} \textbf{324} (2010), 2742--2756.

\bibitem{Nesteerenko:Zaitsev1969}
Zaitsev~G.A.,
Group-invariant study of the sets of limiting geometrie and special Lie subalgebras, 1961,
\textit{Talk thesises of All-USSR Geometric Conference (Kiev, USSR, 1961)}, 1--48 (Russian).


\end{thebibliography}
\end{document}